\documentclass[12pt]{article}
\usepackage{epsfig}

\def \be {\begin{equation}}
\def \ee {\end{equation}}
\def \bea {\begin{eqnarray}}
\def \eea {\end{eqnarray}}

\def \ep{\epsilon}

\def \lab #1 {\label{#1}}

\newcommand{\NP}[1]{Nucl.\ Phys.\ {\bf #1}}
\newcommand{\PL}[1]{Phys.\ Lett.\ {\bf #1}}

\newcommand{\PR}[1]{Phys.\ Rev.\ {\bf #1}}

\newcommand{\hepph}[1]{hep-ph}
\newcommand{\hepth}[1]{hep-th}

\def \as {{\alpha_s}}

\def\ra{\rightarrow}

\begin{document}

\begin{titlepage}
\renewcommand{\thefootnote}{\fnsymbol{footnote}}
\begin{flushright}
YITP-SB-08-38
     \end{flushright}
\par \vspace{10mm}
\begin{center}
{\large \bf
Leading and Non-leading Singularities\\
\vspace{2mm}
in Gauge Theory Hard Scattering
\footnote{Based on talk presented at
{\it Continuous Advances in QCD}; William I.\ Fine Theoretical
Physics Institute, University of Minnesota, May  15-18, 2008.
}}

\end{center}
\par \vspace{2mm}
\begin{center}
{\bf George Sterman}\\
\vspace{5mm}
C.N.\ Yang Institute for Theoretical Physics,
Stony Brook University \\
Stony Brook, New York 11794 -- 3840, U.S.A.\\
\end{center}

\begin{abstract}
This talk reviewed some classic results and recent progress in the resummation of leading and nonleading enhancements in QCD cross sections and of poles in dimensionally-regularized hard-scattering amplitudes.
\end{abstract}
  
  \end{titlepage}

\section{Introduction}

Interest in perturbative methods for QCD and related gauge theories
arises from the phenomenology of 
high energy scattering, and also from the
study of weak-strong duality, as inspired by string theory.
In the following, I'll review some methods and
techniques that have a long history but remain of continuing interest,
along with a few recent advances.   The talk starts
with a perspective on the place of perturbation theory
in an asymptotically free theory,
goes on to recall ideas of factorization and resummation
in perturbative QCD, which leads to a review of one of its classic successes,
so-called $Q_T$ resummation.   It concludes
with applications of
these same ideas to dimensionally-regulated amplitudes
for the scattering of massless partons,
which have been the subject of much recent work.

\section{How We Use Perturbative QCD}

It's worth recalling that despite the early successes of
asymptotic freedom \cite{Gross:1973id,Politzer:1973fx} as a qualitative explanation of scaling,
the applicability of perturbative methods beyond the
parton model was met with a fair amount of skepticism.
The underlying problems, of course, remain with us.   First,
confinement ensures that the quantities we would
most naturally compute in perturbative QCD (pQCD), time-ordered products
of fields,
\bea
\int d^4x\; {\rm e}^{-iq\cdot x}\langle 0|\, T[ \phi_a(x) \dots ]\, |0\rangle\, ,
\eea
 have no $q^2=m^2$ poles for any field (particle) $\phi_a$ that 
transforms nontrivially under color, while
the ``physical"  poles at $q^2=m_\pi^2$, for example, in
\bea
\int d^4x\; {\rm e}^{-iq\cdot x}\langle 0|\, T[ \pi(x) \dots ]\, |0\rangle\, ,
\eea
are not accessible to perturbation theory directly.
And yet we use asymptotic freedom,
up to power-suppressed corrections,
\bea
Q^2\; \hat \sigma_{\rm SD}(Q^2,\mu^2,\alpha_s(\mu)) 
&=&
\sum_n c_n(Q^2/\mu^2)\; \as^n(\mu) + {\cal O}\left({1/Q^p}\right)
\nonumber\\
&=& \sum_n c_n(1)\; \as^n(Q) +   {\cal O}\left({1/Q^p}\right)\, ,
\eea
for single-scale cross sections $\sigma(Q)_{\rm SD}$, so long as they are finite in the zero-mass
limit in perturbation theory, a property known as ``infrared safety".   Various 
total and jet cross sections as well as predictions based
on evolution are of this type, and their phenomenological
successes are well-known.

So, what are we really calculating?  In many cases,
we are computing matrix elements for color singlet currents,
of the general form 
\bea
\int {\rm e}^{-iq\cdot x}\langle 0|\, T[  J(x) J(0)\dots ]\, |0\rangle\, ,
\eea
related to observables by the optical theorem.
Of course, the optical theorem requires a complete sum over final states.
But, in fact, there is another  
class of infrared (IR) safe color singlet matrix elements, related to jets and event shapes,
that have received attention of late.   These matrix elements accompany currents with
the energy-momentum tensor, $T_{\mu\nu}$, schematically,
\cite{Eflow}
\bea
\lim_{R\to \infty} R^2\int dx_0  \int d\hat n\, f(\hat n)\, {\rm e}^{-iq\cdot y}
\langle 0|\, J(0)  T[ \hat n_i\, T_{0i}(x_0,R\hat n) J(y) ]\, |0\rangle\, ,
\eea
with $f(\hat n)$ a ``weight" that controls the contributions of
particles flowing to infinity in different directions, $\hat n$.
With the operator $T_{0i}$ placed at infinity, these matrix elements rather directly represent
the action of a calorimeter.   
If the weight is a smooth function of angles, then even though
the matrix elements for individual final states have IR divergences
in general, they cancel
in sums over collinear 
splitting/merging and
 soft parton  emission, precisely because these rearrangements respect energy flow.
We regularize these divergences dimensionally (typically)
and ``pretend" to calculate the long-distance enhancements
only to cancel them in infrared safe quantities.

\section{Factorization and Resummation}

Beyond the relatively limited class of cross sections that are  directly IR safe,
the predictive power of pQCD depends on factorization \cite{FactPrf,CSS89}.   From factorization
we can derive the evolution familiar from deep-inelastic scattering 
and other single-scale problems, and 
generalizing this viewpoint, we can motivate resummations of enhancements
in multiscale problems.   A factorized cross section takes the general form
\bea
Q^2\sigma_{\rm phys}(Q,m)
=
{\omega_{\rm SD}(Q/\mu,\as(\mu))}\, \otimes\, f_{\rm LD}(\mu,m) + 
{\cal O}\left({1/ Q^p}\right)\, ,
\label{fact}
\eea
where $\mu$ is a  factorization scale, $m$ represents IR scales,
perturbative or nonperturbative, and where $\otimes$ represents
a convolution, typically in parton
fraction or transverse momentum, often accurate to power corrections as shown.
Speculations on new physics are contained  $\omega_{\rm SD}$,
as perturbative (as in SUSY) or nonperturbative (as in
technicolor) extensions of the Standard Model; $f_{\rm LD}$ represents
parton distributions of various sorts, universal among cross
sections sharing the same factorization. 

The familiar ``DGLAP" evolution equations \cite{DGLAP} can be derived from factorization,
just by observing that physical cross sections cannot depend on
the choice of factorization scale
\bea
0=\mu{d\over d\mu} \ln \sigma_{\rm phys}(Q,m)\, ,
\label{muindep}
\eea
which, combined with (\ref{fact}) leads to a separation of variables,
\bea
\mu{d \ln f\over d\mu}= - P(\as(\mu)) = - \mu{d \ln \omega \over d\mu}\, ,
\label{rgfomega}
\eea
where the ``separation constant" $P$ can depend only on the
variables held in common between the short- and long-distance
functions in the factorized expression, $\alpha_s$ and the
convolution variable(s).

The solutions to evolution equations like Eq.\ (\ref{rgfomega}) are
examples of resummation, in this case summarizing  leading 
(and nonleading) logarithms of $Q$,
\bea
\ln \sigma_{\rm phys}(Q,m) \sim \exp\left\{  \int^Q {d\mu'\over \mu'} 
P\left( \alpha_s(\mu')\right) \right\}\, .
\eea
This result is most familiar in the form of DGLAP evolution; as we
shall see, however, its applications are even more wide-ranging.

This sequence of methods and results:  factorization $\ra$ evolution
$\ra$ resummation, varies between observables,
and must be verified for each case.   Such verifications,
or ``factorization proofs" \cite{FactPrf,CSS89,Aybat:2008ct}, rely in general on four features of
gauge theory:
(1)  The operator product expansion, according to which
short-distance dynamics in  $\omega_{\rm SD}$ is incoherent with
long-distance dynamics;
 (2) Jet-jet factorization, or the mutual incoherence of the dynamics
 of particles with
$v_{\rm rel}=c$;
(3) Jet-soft factorization, by which wide angle soft radiation depends only
on the overall color flow in jets \cite{Collins:1981ta,Collins:1981uk};
(4) Dimensionless couplings and renormalizability,
which ensure that infrared singularities are no worse than
logarithmic \cite{S78}.

\section{The Classic Case: $Q_T$ Resummation}

What makes factorization necessary, and
evolution and resummation so rewarding, is that
every final state from a hard scattering carries the imprint 
of QCD dynamics from  all distance scales.
We will illustrate how these ideas play out in
 {\it the} classic application of resummed pQCD, the transverse
 momentum distribution for Dell-Yan pairs \cite{Collins:1981uk,Collins:1984kg}.
 
\begin{figure}[h]
\begin{center}
\epsfxsize=5cm \epsffile{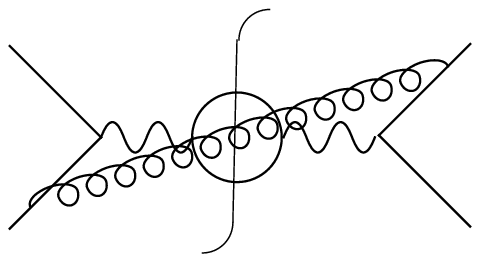} \qquad \epsfxsize=5cm \epsffile{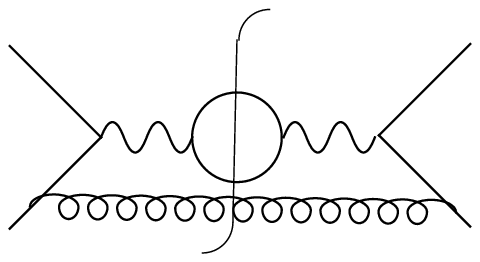}
\caption{LO gluon emission diagrams for $\hat\sigma$, Eq.\ (\ref{LOhatsig}).\label{LOfig}}
\end{center}\end{figure}

We start with the transverse momentum distribution at order $\as$
for the purely partonic process  
\begin{equation}
     q(p_1) + \bar{q}(p_2) \rightarrow \gamma^*(Q)+g(k)\, .
\end{equation}
At lowest order (LO), ${\bf k} = - {\bf Q}_T$, and the partonic
cross section is free of infrared divergences.   
The corresponding factorized expression for the LO {\it hadronic} cross
section is 
	\bea
{d\sigma_{{NN}\ra \mu^+\mu^-+X} (Q,p_1,p_2) \over dQ^2 d^2{\bf Q}_T }
&=& 
\int_{\xi_1,\xi_2}\ \sum_{a={ q\bar{q}}}
 { {d\hat{\sigma}_{a\bar{a} \ra \mu^+\mu^-(Q)+X }(Q,\mu,\xi_ip_i,{\bf Q}_T)
 \over dQ^2 d^2{\bf Q}_T }} 
\nonumber \\
&\ & \hspace{10mm}  \times \ {f_{a/N}(\xi_1,\mu)\, f_{\bar{a}/N}(\xi_2,\mu)}\, .
\label{DYQTfact}
\eea
The LO diagrams for the measured-$Q_T$ cross section are shown in
Fig.\ \ref{LOfig},
where the short-distance factor (the analog of $\omega_{\rm SD}$ above) is
\begin{eqnarray}
&& 
     \frac{d\hat{\sigma}_{q\bar{q}\ra \gamma^*g}^{(1)} }  
{dQ^2\,d^2{\bf Q}_T} =
\sigma_0 \frac{\alpha_s C_F}{\pi^2}
\left(1-\frac{4 {\bf Q}_T^2}{(1-z)^2 \xi_1\xi_2S}\right)^{-1/2}
\nonumber\\
\ \nonumber \\
&& \hspace{40mm} \times 
\;\left[\frac{1}{{\bf Q}_T^2}  \frac{1+z^2}{1-z} -\frac{2z}{(1-z)Q^2}\right] \, ,
\label{LOhatsig}
\end{eqnarray}
with $\sigma_0$ the LO total cross section,
This expression, and the
corresponding factorized cross section (\ref{DYQTfact}),
 is well-defined as long as ${\bf Q}_T \ne 0$ and $z = Q^2/\xi_1\xi_2S \ne 1$.

Now the leading behavior for ${\bf Q}_T\ll Q$ can be found
by considering the $z$ integral.    When  $Q^2/S$ is not too close to unity,
 the phase space factor in (\ref{LOhatsig})
and the parton distribution functions (PDFs)
 can be treated as nearly constant over the
physical range of $z$, which then gives a logarithmic integral,
\bea
\frac{1}{{\bf Q}_T^2}\; \int^{1-2|{\bf Q}_T|/Q}_{Q^2/S}\frac{dz}{1-z} 
\sim
\frac{1}{{\bf Q}_T^2}\;  \ln\left[\, \frac{Q}{|{\bf Q}_T|}\, \right]\, .
\eea
This approximation gives a neat prediction for $Q_T$ dependence at fixed $Q$,
	\bea
{d\sigma_{{NN}\ra \mu^+\mu^-+X} (Q,{\bf Q}_T) \over dQ^2 d^2{\bf Q}_T}
&\sim& \frac{\alpha_s C_F}{\pi}\  \frac{1}{{\bf Q}_T^2}\;  \ln\left[\, \frac{Q}{|{\bf Q}_T|}\, \right]
\nonumber \\
&\ &  \hspace{-40mm} \times\ \sum_{a=q\bar{q}}
\int_{\xi_1\xi_2}  \frac{d\hat{\sigma}_{a\bar{a}\ra \mu^+\mu^-(Q)+X}(Q,\mu)}{dQ^2} 
\ f_{a/N}(\xi_1,\mu)\, f_{\bar{a}/N}(\xi_2,\mu)\, ,
\eea
which we can
 compare, for example,  to the
 transverse momentum of the Z boson at the Tevatron.
\begin{figure}[h]
\begin{center}
\epsfxsize=9cm \epsffile{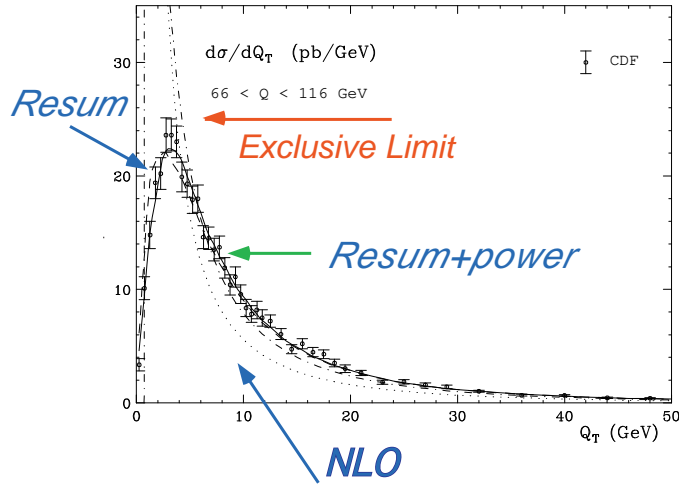}
\caption{Transverse momentum dependence of the Z boson as 
observed by the CDF collaboration.   \label{Zpt}}
\end{center}
\end{figure}
As can be seen from Fig.\ \ref{Zpt}, taken from Ref.\ \cite{Kulesza:2002rh},  a simple
 $\ln Q/Q_T$-dependence works pretty well for ``large" $Q_T$,
 less than but of the order of $Q=m_Z$, but at smaller $Q_T$ 
 the distribution reaches a maximum, then  decreases
near the ``exclusive" limit, at $Q_T=0$, corresponding to parton model kinematics.
 Indeed, most events are at  ``low" $Q_T \ll m_Z$, where the LO
 cross section diverges.
 To understand the distribution in this range, we turn to transverse
 momentum resummation, which, as we shall see, controls 
logarithms of $Q_T$ to all orders in $\alpha_s$.
As suggested above, we can resum logarithms of $Q_T$
by developing variant factorizations and separations of variables.

In brief, the factorization we will exhibit reflects a relatively simple physical picture.
The active quark and antiquark   arrive at 
the point of annihilation with nonzero transverse momenta, due to
gluons radiated in the transition from the initial state.
Now before the collision, the quark
and antiquark radiate independently, reflecting a lack of
overlap between their Coulomb fields.
Similarly, after the collision, 
final-state radiation occurs too late to affect the cross section, that is,
the net probability of annihilation into an electroweak vector boson
with a given $Q_T$.
These considerations are summarized by  $Q_T$-factorization,
in the form \cite{Collins:1984kg}
\bea
  &\ &   {d\sigma_{NN\rightarrow QX} \over dQ^2 d^2Q_T}
=  \int d\xi_1 d\xi_2\ d^2{\bf k}_{1T}d^2{\bf k}_{2T}d^2{\bf k}_{sT}\,
{\delta^2\left ( {\bf Q}_T-{\bf k}_{1T}-{\bf k}_{2T}-{\bf k}_{sT}\right)}\
   \nonumber\\
   &\ & 
 \ \hspace{5mm} \times \sum_{a=q\bar q}\; {H_{a \bar a}(\xi_1p_1,\xi_2p_2,Q,}{ n})_{a\bar{a}\rightarrow Q+X } 
\nonumber   \\
 &\ &  \hspace{5mm} \times\ {{\cal P}_{a/N}}(\xi_1, {p_1\cdot n,{\bf k}_{1T}})\, 
{ {\cal P}_{\bar{a}/N}}(\xi_2,{p_2\cdot n,{\bf k}_{2T} })\  U_{a \bar{a}}({{\bf k}_{sT},n})\, .
\label{QTfact}
\eea
Here the ${\cal P}'s$ are new {\it transverse momentum-dependent} PDFs,
and in the general case we also need a new function labelled,
$U$, a {\it soft function} that describes wide-angle radiation.
 Symbolically, in the spirit of the general factorization, Eq.\ (\ref{fact}), we can write 
\bea
  {d\sigma_{NN\rightarrow QX} \over dQ^2 d^2Q_T}
&=&
\sum_{a=q\bar q}\; H_{a \bar a} \times {{\cal P}_{a/N}}(\xi_1, {p_1\cdot n,k_{1T}})\, 
{ {\cal P}_{\bar{a}/N}} (\xi_2,{p_2\cdot n,k_{2T} })\nonumber
\\
&\ &  \hspace{5mm} \otimes_{\xi_i,k_{iT}}\  U_{a \bar{a}}({k_{sT},n})\, .
\label{QTfactshort}
\eea
What we are going to do is {\it derive} the $k_T$ dependence of the ${\cal P}$'s
from this relation.
For the purposes of this talk, we proceed intuitively and
with broad strokes; much more careful analyses
can be found in \cite{Collins:1981uk,Collins:1984kg}, and \cite{Contopanagos:1996nh}.

In Eq.\ (\ref{QTfact}) we encounter
new invariants, $p_i\cdot n$, formed from a fixed vector {$n^\mu$}.
We can think of $n^\mu$ as being used to apportion 
real and virtual gluons of momentum $k$ into the various factors
in (\ref{QTfact}), according to the scheme:
\bea
p_a\cdot k < n\cdot k \ &\Rightarrow&\ k\ \in {\cal P}_a \nonumber \\
p_a\cdot k,\; p_{\bar{a}}\cdot k > n\cdot k\ &\Rightarrow&\ k \in U\, .
\eea
It is the variables $p_a\cdot n$ that will play the role of 
factorization scales.
Before reviewing this analysis, 
we go to impact parameter space, replacing the convolution in $k_{i,T}$
by a product after the Fourier transform with
${\rm  e}^{i\vec{Q}_T\cdot \vec{b}}$, giving, in place of (\ref{QTfact}),
\bea 
  {d\sigma_{NN\rightarrow QX}(Q,b) \over dQ^2 }
&=&  \int d\xi_1 d\xi_2 
  \ {H(\xi_1p_1,\xi_2p_2,Q,}{ n})_{a\bar{a}\rightarrow Q+X } 
 \nonumber  \\
 &\ & 
\hspace{-5mm} \times {{\cal P}_{a/N}}(\xi_1, {p_1\cdot n,b})\, 
{ {\cal P}_{\bar{a}/N}}(\xi_2,{p_2\cdot n,b })\  U_{a \bar{a}}({b,n})\, .
\label{impactfact}
\eea
We are now ready once again to resum by separating variables.

The physical impact parameter cross section of Eq.\ (\ref{impactfact})
is independent of both $\mu_{\rm ren}$ and of the vector $n^\mu$.   
As a result, we have two equations that express this independence,
\bea
 \mu_{\rm ren}{d\sigma \over  d\mu_{\rm ren}} = 0\, , \hspace{10mm} n^\alpha{d\sigma \over dn^\alpha}=0\, .
\label{2indep}
\eea
These equations represent the scale variation and the 
boost invariance of the theory.
The solutions to pairs of equations of this kind were developed in this 
context by Collins and Soper \cite{Collins:1981uk} and by Sen \cite{Sen:1981sd}.

 Now variations from the jets must cancel variations
 from the short-distance function $H$ and from the soft function $U$,
 which depend on different variables.   
 This analysis gives
\bea
p\cdot n\, \frac{\partial}{\partial p\cdot n}\, \ln\, {\cal P}(p\cdot n/\mu,b\mu) 
= \frac{1}{2} G(p\cdot n/\mu,\as(\mu)) + \frac{1}{2} K(b\mu,\as(\mu)),
\label{PGK}
\eea
where $G$ matches $H$, and $K$ matches $U$.  
On the other hand, renormalization is independent  of $n^\mu$, which implies 
\bea
\mu\, \frac{\partial}{\partial \mu} \left[\, G(p\cdot n/\mu,\as(\mu)) + K(b\mu,\as(\mu))\, \right] = 0\, ,
\eea
from which we find
\bea
\mu\, \frac{\partial}{\partial \mu} \, G(p\cdot n/\mu,\as(\mu))\
=\ \gamma_K(\alpha_s(\mu))\ =\ -\
\mu\, \frac{\partial}{\partial \mu}\, 
K(b\mu,\as(\mu))\, .
\label{GAK}
\eea
It is the combination of Eqs.\ (\ref{PGK}) and (\ref{GAK}) that gives the basic results.

We solve Eq.\ (\ref{GAK}) first, 
\bea
G(p\cdot n/\mu,\as(\mu)) + K(b\mu,\as(\mu)) &=& G(1,\as(p\cdot n)) + K(1,\as(1/b))
\nonumber
\\
&\ & 
\hspace{3mm} -\ \int_{1/b}^{p\cdot n} \frac{d\mu'}{\mu'}\, \gamma_K(\alpha_s(\mu'))\, .
\eea
Inserting this result in the consistency equation (\ref{PGK}) for the jet enables
us to integrate $p\cdot n$ and get double logs in $b$,
 which, when inverted back to $Q_T$ space,
 produce the leading behavior $\alpha_s^n\frac{\ln^{2n-1}(Q/Q_T)}{Q_T}$
 at each order in $\alpha_s$, along with nonleading contributions (which require an
 analysis of the soft function $U$).
 When carried out in detail (with attention paid to 
 nonperturbative corrections from large $b$), this approach 
 can describe the data of Fig.\ \ref{Zpt}
 all the way to $Q_T=0$. \cite{Kulesza:2002rh,DYQtrefs}
The resulting expression can be summarized as
\bea
\label{crsec}
       \frac{d\sigma_{NN{\rm res}}}{dQ^2\,d^2 \vec Q_T}
       &=&   
{\sum_a}\, 
{    \int \frac{d^2b}{(2\pi )^2} \, e^{i\vec{Q}_T\cdot \vec{b}}}\, 
 {\exp\left[ \,E_{a\bar{a}}^{\rm PT} (b,Q,\mu)\,\right]} 
\nonumber\\
&\ & \hspace{-12mm} \times\ {\sum_{a=q\bar{q}}
\int_{\xi_1\xi_2} {H(\xi_1p_1,\xi_2p_2,Q,}{ n})_{a\bar{a}\rightarrow Q+X }
\ f_{a/N}(\xi_1,1/b)\, f_{\bar{a}/N}(\xi_2,1/b)} \, ,\nonumber\\
\eea
with a ``Sudakov" exponent  that, as anticipated, links large and low virtuality,
\bea
E_{q\bar{q}}^{\rm PT} = -
\int_{1/b^2}^{Q^2} {d \mu^2 \over \mu^2} \;
\left[ 2A_q(\as(\mu))\,
\ln\left( {Q^2 \over \mu^2} \right)\, + 2B_q(\as(\mu))\right] \, ,
\eea
where \cite{Collins:1984kg} $B_q$ is related to $(K+G)_{\mu=p\cdot n}$ and at
lowest order $A_q=\gamma_K/2$, and where the lower limit $1/b$ of the integral
in the exponent
generates the leading logarithmic $Q_T$ dependence.

\section{Poles in Color Exchange Amplitudes}

Color exchange is a feature
central to the analysis of hard scattering and 
jet and heavy particle production at hadron colliders.
As intermediate results in calculations of short-distance functions,
and as a subject of interest in their own right,
 multiloop scattering amplitudes 
in dimensional regularization have received considerable
attention \cite{Magnea:1990zb,Catani:1998bh,Sterman:2002qn,MertAybat:2006mz}.
We conclude with a sketch of how the general methods
described above lead to important results for these ampliudes.

We consider a partonic process, denoted:
${\rm f}:\quad f_A(p_A,r_A) + f_B(p_B,r_B) \to f_1(p_1,r_1) +
f_2(p_2,r_2) + \dots $, where we restrict ourselves to
wide-angle scattering.
The amplitude for any such process can be expanded in
a basis of color tensors $c_L$ linking the external partons,
\begin{eqnarray}
&& 
\label{amp}
{\cal M}^{\rm[f]}_{\{r_i\}}\left(p_j ,\frac{Q^2}{\mu^2},\as(\mu^2),\ep \right)
=
{\cal M}^{\rm[f]}_{L}\left( p_j,\frac{Q^2}{\mu^2},\as(\mu^2),\ep \right)
\, \left(c_L\right)_{\{r_i\}}\, ,
\end{eqnarray}
with infrared singularities regularized by going to $4-2\epsilon$
dimensions with $\epsilon < 0$, after renormalization has been performed.
Examples of the $c_I$s are singlet and octet exchange in the $s$-channel
of quark-antiquark scattering.
 We need to control poles in $\epsilon$ for factorized calculations at 
fixed order, and, for resummation, to all orders.

\begin{figure}[t]
\begin{center}
 \epsfxsize=10cm \epsffile{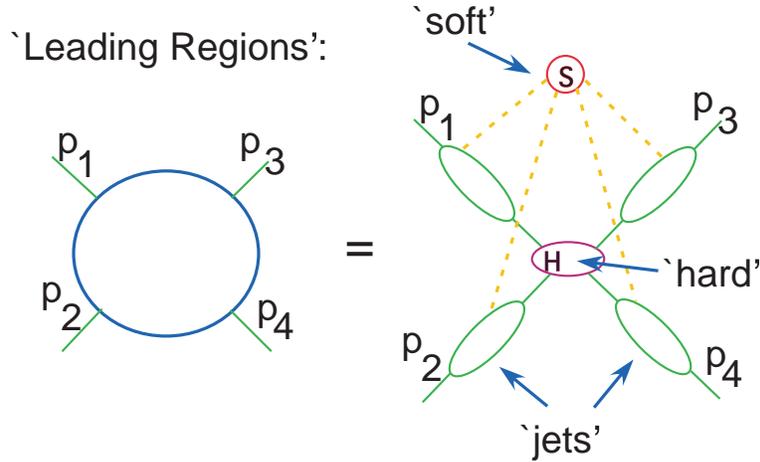}
 \caption{Leading regions for $2\ra 2$ scattering. \label{lrs}}
\end{center}
\end{figure}

\begin{figure}[t]
\begin{center}
 \epsfxsize=10cm \epsffile{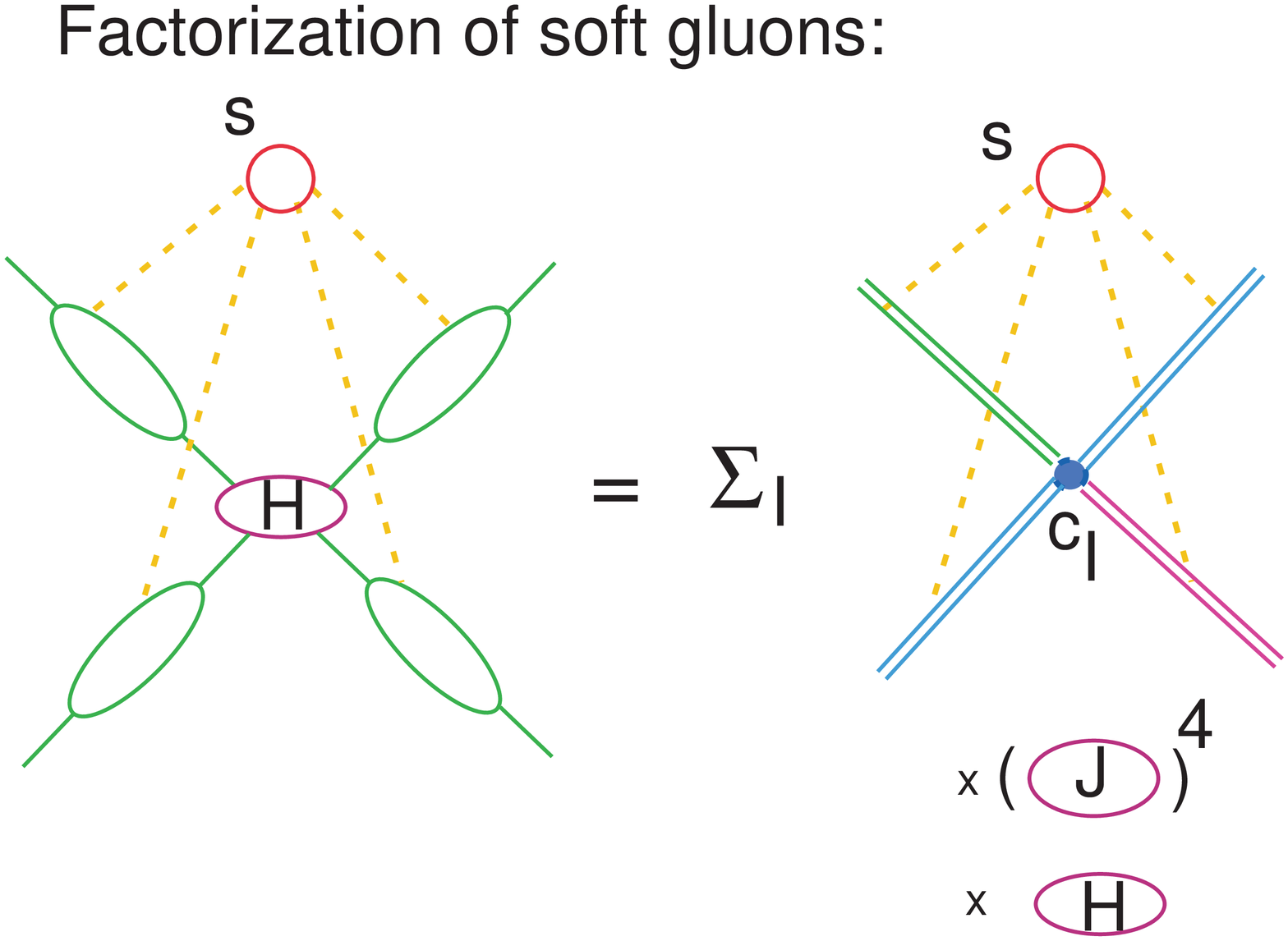}
 \caption{Soft-jet factorization for wide-angle scattering. \label{wifact}}
 \end{center}
\end{figure}

Double logs and poles in dimensional regularization
 are associated with {\it leading regions} \cite{S78,Akhoury:1978vq} in the
loop momentum space for arbitrary graphical contributions to the 
amplitude.  These take the general form shown in Fig.\ \ref{lrs}.
Leading regions are characterized by {\it jet subdiagrams}, consisting of
lines parallel to the external momenta $p_i$, a {\it short-distance
subdiagram} ($H$), with only lines off-shell
by order of the momentum transfer(s), and a {\it soft subdiagram} ($S$) with lines
whose momenta vanish.    
Historically, it was the soft subdiagram that was seen as
a problem for the control of infrared behavior. \cite{Callan:1974zy}
In Fig.\ \ref{lrs}, however, we encounter the same 
cast of characters as for the $Q_T$ analysis in Drell-Yan,
 and the analogous factorization is shown schematically in Fig.\ \ref{wifact},
in which jet-soft factorization
 separates jet and soft dynamics in this more complex scattering process.
 \cite{Sen:1982bt}
For the amplitude, the IR regularization variable
 $\epsilon = 2 - d/2$ plays the role of $b$ in the $Q_T$ cross section.
 
In summary, we can write a factorized expression for ${\cal M}$,
\begin{eqnarray}
&&
\label{facamp}
{\cal M}^{\rm[f]}_{L}\left(p_i,\frac{Q^2}{\mu^2},\as(\mu^2),\ep \right)
= \prod_{f=A,B,1,2} 
{J_f^{\rm[virt]}\left(\frac{Q^2}{\mu^2},\as(\mu^2),\ep \right)}
\nonumber
\\
&& \hspace{15mm}\times
{{\bf S}^{\rm[f]}_{LI}\left(p_i,\frac{Q^2}{\mu^2},\as(\mu^2),\ep \right)} \; h^{\rm[f]}_{I}\left(\wp_i,\frac{Q^2}{\mu^2},\as(\mu^2)
\right)\, ,
\label{MJSH}
\end{eqnarray}
where the jet functions $J_f $ for parton $f$ can be identified 
with the square roots of the corresponding
singlet form factors, $\sqrt{\Gamma_{\rm singlet}^f(Q^2)}$
\cite{Sterman:2002qn},
the soft functions are matrices labelled by color exchange  (singlet, octet \dots),
and all factors require dimensional regularization.
We return to the soft function ${\bf S}^{\rm[f]}$ below.
   
The same analysis as for Drell-Yan $Q_T$
described above, starting with factorization and arriving at resummation,
 gives the following explicit expression \cite{Magnea:1990zb} modeled
on the work of Collins and Soper \cite{Collins:1981uk} and of Sen \cite{Sen:1981sd}:
\begin{eqnarray}
 & & \hspace{-1cm} \Gamma \left( \frac{Q^2}{\mu^2}, \as(\mu^2), 
  \epsilon \right) ~=~ \exp \left\{ \frac{1}{2} \int_0^{- Q^2} 
  \frac{d \xi^2}{\xi^2} \Biggl[
  {\cal K} \left(\epsilon, \as(\mu^2) \right) \right. 
  \label{sol} \\ 
  &\ & \hspace{5mm} +\ \left. {\cal G} \left(-1, \as
  \left(\xi^2,\epsilon \right), \epsilon \right) 
  + \frac{1}{2} \int_{\xi^2}^{\mu^2} 
  \frac{d \lambda^2}{\lambda^2} \gamma_K \left(\as
  \left(\lambda^2,\epsilon \right) \right) \Biggr] 
  \right\}~, \nonumber
\eea
where the running coupling is treated as $\epsilon$-dependent.
All levels of exponentiating poles are generated by the
anomalous dimensions ${\cal G}$, ${\cal K}$ and $\gamma_K=-\mu d{\cal K}/d\mu$.
(The functions are
${\cal G}$ and ${\cal K}$ are related to, but not identical with the analogous
functions above.)
  The relations of such QCD results to supersymmetric Yang-Mills theories
  were explored in several talks at this workshop
   (see also the recent review
  by Alday and Roiban \cite{Alday:2008yw}).
Double poles are generated from
$\gamma_K$, which is familiar as
the so-called ``cusp" anomalous dimension \cite{Korchemsky:1987wg}.
 A complete portrait of single poles at each order requires 
 the $\epsilon$-dependent function ${\cal G}$ \cite{Dixon:2008gr},
which  also generates finite coefficient functions in  $\Gamma_{\rm singlet}$
\cite{Moch:2005tm}.   To find a field-theoretic interpretation for ${\cal G}$, we once
again turn to the factorization approach, this time for the singlet
form factor itself, as illustrated in Fig.\ \ref{singfact}.
\begin{figure}[t]
\begin{center}
\hbox{\hskip 0.8 in \epsfxsize=10cm \epsffile{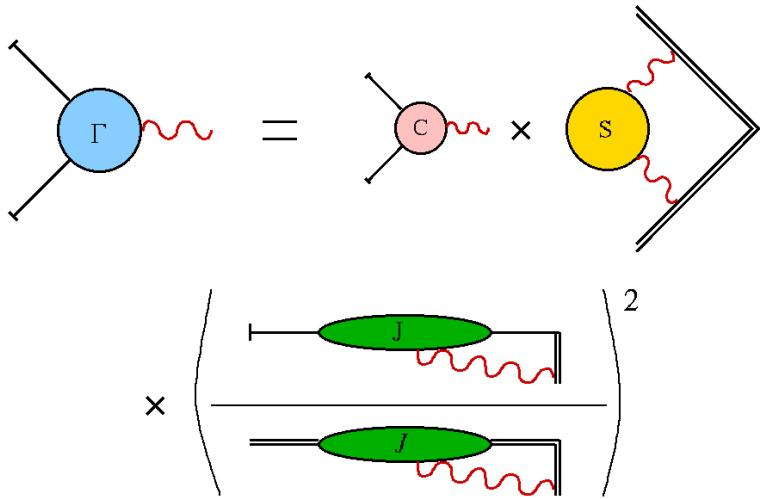}}
\vskip 0.2 in
\caption{Factorization of the singlet form factor. From Ref.\ 
{\protect \cite{Dixon:2008gr}}. \label{singfact}}
\end{center}
\end{figure} 
In the figure, soft radiation is organized in a singlet product of 
light-like Wilson lines,
\bea
  {\cal S} \left(  \as (\mu^2), \epsilon \right) =
  \langle 0 | \Phi_{\beta_2} (\infty,0) \, \Phi_{\beta_1} (0, - \infty) 
  \, | 0 \rangle\, ,
\eea
where $\Phi_\beta(\infty,0) \equiv
 P\, \exp[ -ig\int_0^\infty d\lambda \beta\cdot A(\lambda\beta)]$, and where
 we may take $\beta_1\cdot \beta_2=1$.
Such an expectation value obeys \cite{Korchemsky:1992xv}
\begin{eqnarray}
  \mu \frac{d}{d \mu} \log{{\cal S}} \left( 
  \as (\mu^2),\varepsilon \right) & = &  G_{\rm eik} \left( 
  \as (\mu^2) \right) - \frac{1}{2} \int_0^{\mu^2} \frac{d \xi^2}{\xi^2} 
  \gamma_K \left( \overline{\alpha} (\xi^2, \varepsilon) \right)\, ,
\end{eqnarray}
in terms of the same $\gamma_K$
and a new anomalous dimension
 $G_{\rm eik}$ that organizes non-collinear poles.

Following this analysis, the full ${\cal G}$ for the form factor 
in Eq.\ (\ref{sol}) can be written as \cite{Dixon:2008gr}
\begin{eqnarray}
{\cal G} = 2B+ G_{\rm eik} + \beta(g) \frac{\partial}{\partial g}\, C(\alpha_s(Q))\, ,
\end{eqnarray}
with $B$ the $N$-independent coefficient in spin-$N$ leading-twist
operators for parton $i$, and with
$C$ the short-distance function shown in Fig.\ \ref{singfact}.
 Similar combinations have been encountered in analyses of deep-inelastic scattering and Drell Yan in
 Refs.\ \cite{Idilbi:2006dg,Becher:2006mr,Becher:2007ty}.

The remainder of the dimensional dependence in the general
amplitude, Eq.\ (\ref{MJSH}) is generated by a matrix of anomalous
dimensions for the soft functions
\cite{colorflow,Sterman:2002qn}
\begin{eqnarray}
\label{expoS}
&& 
{{\bf S}^{\rm[f]}\left(\frac{Q^2}{\mu^2},\as(\mu^2),\ep \right)} =
{\rm P}~{\rm exp}\left[ -\frac{1}{2}\int_{0}^{Q^2} \frac{d\tilde{\mu}^2}{\tilde{\mu}^2}
{\bf \Gamma}_S^{\rm[f]} \left(\as\left(\tilde{\mu}^2,\ep\right)\right) \right]\, .
\nonumber\\
\end{eqnarray}
The one-loop expressions for arbitrary ${\bf \Gamma}^{\rm[f]}_S$ were computed in
\cite{colorflow}, and the two-loop expressions  in \cite{MertAybat:2006mz}.
Remarkably, the one- and two-loop contributions
are proportional \cite{MertAybat:2006mz},
\bea
&&{{\bf \Gamma}_S = \frac{\as}{\pi}\, \left( 1 + \frac{\as}{2\pi}\, K\right) \; 
{\bf \Gamma}_{S}^{(1)}}\ +\
\cdots\ ,
\eea
with
with the same constant, $K=C_A(67/18-\zeta_2) -(5/9)n_f$, that appears 
in $\gamma_K$ for parton $i$,
\bea
\gamma_K = 2C_i\, \frac{\as}{\pi}\, \left( 1 + \frac{\as}{2\pi}\, K\right) + \dots \;
\eea
This suggests an exact one-loop anomalous dimension,
supplemented by a ``CMW" scheme for $\as$ \cite{Catani:1990rr}.
If this conjecture turns out to hold, there is a
deep simplicity inherent in infrared vector exchange, even in QCD.

\section{Summary}

I have shown how the factorization properties
of gauge theories serve as keys to resummation.   For double-logarithmic, or ``Sudakov"
corrections, resummation follows from 
two equations, one associated with boost invariance, and another with scale 
variations (scale invariance for conformal theories).
 The basic factorization structure and its consequences
 are not limited to weak coupling.
Whether at weak or strong coupling, many of the
the long-distance properties of gauge theories
can be organized 
quite explicitly in both cross sections and the perturbative S-matrix.

\section*{Acknowledgments}
I thank the organizers for an invitation, and for support in
the course of this stimulating workshop.  I also thank my
collaborators in the recent work reported on here, 
Mert Aybat, Lance Dixon and Lorenzo Magnea.
This work was supported in part by
by the National Science Foundation,  
grants PHY-0354776, PHY-0354822 and PHY-0653342.

\end{document}